\documentclass[bibyear]{aa}

\usepackage{graphicx}
\usepackage{txfonts}
\usepackage{bm}
\usepackage{float}
\usepackage{enumerate}
\usepackage{natbib}
\usepackage{hyperref}
\usepackage{xcolor}

\begin{document} 

   \title{Data-calibrated point spread function prediction}
   \subtitle{General description of the method and demonstration on MUSE-NFM}

    \author{
        Arseniy Kuznetsov \inst{1,2,3}
        \and
        Benoit Neichel \inst{2}
        \and
        Sylvain Oberti \inst{1}
        \and
        Thierry Fusco \inst{2,3}
    }
    \institute{
        European Southern Observatory, Karl-Schwarzschild-str-2, 85748 Garching, Germany\\
        \email{akuznets@eso.org}
        \and
        Aix Marseille Univ, CNRS, CNES, LAM, 13013 Marseille, France
        \and
        DOTA, ONERA, Universit\'e Paris Saclay, F-91123 Palaiseau, France
    }

   \date{Received 17 September 2025 / accepted 19 February 2026}

\abstract
{
    Precise knowledge of the point spread function (PSF) underpins many data analysis steps in astronomy, from photometry and astrometry to source de-blending and deconvolution. In adaptive optics (AO) observations, however, the PSF is highly variable with wavelength, field position, and observing conditions, making it difficult to model. Traditional PSF reconstruction (PSF-R) requires full AO telemetry and complex infrastructures, limiting its routine use, especially for tomographic systems.
    
}
{
    We present a practical framework for fast, accurate, and data-calibrated PSF modeling that captures the spatial and spectral variability of AO-corrected PSFs without relying on complete AO telemetry.
}
{
    Our approach builds on a Fourier-based PSF model inspired by astro-TIPTOP. As inputs, our model uses only a compact set of physically meaningful parameters retrievable from the ESO archive. A lightweight neural network corrects these inputs to achieve the best match with real data. It is trained end to end with the PSF model, allowing it to learn any miscalibrations directly from on-sky data.
}
{
    The framework achieves high accuracy on on-sky data. On a test set of MUSE-NFM standard stars, it yields median errors of 13.5\% in the Strehl ratio and 10.9\% in the core full width at half maximum (FWHM). In crowded MUSE-NFM observations of $\omega$ Centauri, the method predicts dozens of off-axis, wavelength-dependent PSFs with a Strehl error of <5\% and a FWHM error of 4.6\%, enabling source separation without per-star PSF extraction.
}
{
    Our compact, physics-informed, and data-calibrated model delivers accurate, polychromatic, and field-varying PSFs without relying on full AO telemetry. While demonstrated on MUSE-NFM, the method is still transferable to other AO-assisted instruments.
 }

\keywords{
    instrumentation: adaptive optics --
    instrumentation: spectrographs --
    point spread function --
    methods: data analysis --
    methods: numerical --
    telescopes: VLT
}

\maketitle

\section{Introduction}

In astronomical imaging, the accuracy of scientific analysis depends strongly on the point spread function (PSF). A broadened PSF reduces the clarity of information that can be extracted from observations. For ground-based telescopes, atmospheric turbulence is the main cause of PSF degradation. Adaptive optics (AO) systems counteract this by partially correcting atmospheric turbulence-induced wavefront errors, pushing the attainable angular resolution toward the diffraction limit.
However, AO correction is inherently imperfect due to the engineering limitations of the real systems, leaving residual wavefront errors (AO residuals) that prevent the PSF from reaching the diffraction limit. Consequently, in many astronomical applications, the accuracy of scientific analysis remains strongly constrained by PSF quality, but can be greatly enhanced through precise knowledge of the PSF structure, which allows it to be disentangled from the astrophysical signal.
For example, when observing the Galactic center, the precise centroiding of stars facilitated by an accurate PSF model can improve the precision for kinematic measurements and thus help one study the interaction of Sgr A$^{*}$ with the surrounding stellar environment. However, the PSF estimation error of only a few percent can induce astrometric biases that exceed the signal \citep{Neichel:14}, necessitating the accurate characterization of the PSF morphology.
Similarly, when studying the stellar kinematics within the dense cores of globular clusters, the precise tracking of stellar trajectories can reveal signatures of intermediate‐mass black holes. However, attaining radial‐velocity uncertainties below 3 km/s requires PSF full width at half maximum (FWHM) errors of under 2\% \citep{Kamann:13}. In such crowded environments, detailed PSF characterization is crucial for de-blending adjacent sources and thereby enhancing both astrometric and photometric accuracy.
Extragalactic studies require one to resolve morphological and spectral details to probe internal dynamics, yet errors of over 20\% in the PSF FWHM can mimic a galaxy's inclination and bias rotation velocity measurements \citep{Bouche:15}. In integral‐field spectroscopy (IFS), imperfect knowledge of PSF elongation can distort metallicity gradients \citep{Contini:12}. Similarly, imperfect PSF shape knowledge can obfuscate the true shape of compact star-forming clumps, skewing spatially resolved star-formation-rate measurements \citep{Cibinel:15}. Hence, accurately characterizing the PSF's variation over wavelength and field position is vital to separate it from true astrophysical features.
When talking about Solar System observations, PSF deconvolution markedly enhances the contrast of surface features on extended bodies \citep{Fetick_gap:19}. To ensure reliable restoration, however, PSF shape errors must be below 5\% \citep{Lau:23}, as accurate knowledge of the PSF structure can regularize the inherently ill-posed deconvolution problem.

To summarize this discussion, a well-characterized shape of long-exposure PSF can be instrumental in various post-processing applications, including precise photometry and astrometry, source subtraction, deconvolution, and more. The comprehensive list of PSF accuracy requirements for different applications can be found in \citet{Neichel:24}, Fig.~9.

A direct approach to characterizing the PSF is to extract it from science exposures using point sources across the field. In practice, however, this is rarely feasible since AO-corrected long-exposure PSFs usually exhibit complex structures that vary significantly with field position, wavelength, and atmospheric conditions. On ELT-class telescopes, for which advanced AO systems will be essential, the PSF complexity will be even greater. Consequently, reference PSFs derived from field reference stars can de-correlate significantly from the science target's PSF, limiting their usefulness for accurate post-processing \citep{Fetick_PSFAO:19}. The problem is compounded by the frequent absence of isolated point sources, particularly in narrow-field instruments such as MUSE-NFM \citep{Leibundgut:19} and HARMONI \citep{Thatte:22}, or in deep extragalactic fields.

In summary, advanced post-processing of AO-assisted observation data requires a PSF-modeling framework that captures the full spatial and spectral variability of long-exposure PSFs across the field. In addition, it needs to predict the PSF morphology even in regions without isolated point sources.

\section{Comparing approaches to PSF modeling}
\subsection{PSF reconstruction}

To address this challenge, several techniques can be used \citep[see][for a comprehensive overview]{BeltramoMartin_review:20}. Among these, PSF reconstruction (PSF-R) is particularly prominent. Proposed in \citet{Veran:97}, PSF-R exploits AO telemetry, specifically long sequences of wavefront sensor (WFS) measurements and deformable-mirror (DM) commands, to reconstruct the long-exposure AO-corrected PSF. However, despite several demonstrated successful applications \citep[see][]{Harder:00, Martin:16, Massari:20, BeltramoMartin:19}, systematic on-sky deployment of PSF-R has remained elusive over the past three decades, with practical implementations largely confined to experiments on single-conjugate AO (SCAO) systems. This limitation mainly arises from several practical hurdles.
\begin{itemize}
    \item One of the key practical limitations is the lack of systematic access to raw AO telemetry required for PSF-R. Archiving and transmitting these data require specialized data infrastructure and large storage capacities, which are not feasible on all existing systems. For example, while Keck Observatory routinely records AO telemetry \citep{Ragland:16} for PSF-R needs, ESO's VLT stores such data only intermittently. The issue of the required volume will be exacerbated on ELT-class systems, for which the sheer scale of AO systems will generate significantly larger telemetry volumes.
     
    \item Calibrating PSF-R is another major challenge. For example, PSF-R algorithms are sensitive to errors in determining the $C_n^2$ profile and WFS gains \citep[see][]{BeltramoMartin:19}, necessitating calibration to achieve the best match to on-sky data. Furthermore, since PSF-R depends exclusively on WFS measurements, it cannot recover any phase aberrations that the AO system does not directly sense. Therefore, contributors such as high-order uncorrected atmospheric aberrations, phasing errors on a segmented primary, non-common path aberrations (NCPAs), and the low-wind effect must be incorporated via external simulation or measurement and carefully adjusted against real data to reproduce observed PSFs. Although techniques such as PRIME \citep{BeltramoMartin:19} have been proposed to facilitate PSF-R calibration, the process remains labor- and computationally intensive. Moreover, a robust and generalizable PSF-R calibration requires an extensive on-sky dataset that spans various atmospheric conditions and AO-system operating regimes. In practice, however, the limited volume of archived AO telemetry impedes the collection of such a representative dataset. Thus, it forces the initiation of dedicated data-collection campaigns, which are complicated by the scarcity of available observing time on operational telescopes.

    \item Point spread function reconstruction for tomographic AO is significantly more challenging than for SCAO. Although theoretical extensions of PSF-R exist for tomographic AO \citep{Wagner:22}, the required algorithms and numerical routines are far more complex and computationally demanding. Moreover, tomographic systems that employ multiple WFSs and deformable mirrors generate large volumes of telemetry, which brings us back to the first point. Future instruments such as MAVIS on the VLT and HARMONI/MORFEO on the ELT will rely heavily on this type of AO correction.
    
\end{itemize}

Within this framework, the primary objective of this paper is to introduce a robust and practical PSF-modeling strategy that overcomes the listed limitations of conventional PSF-R. Our approach is designed to be data-efficient, computationally lightweight, applicable across diverse AO regimes, and readily calibratable with existing on-sky data.
This article presents the general methodology and validates it on ESO's MUSE instrument in the laser tomography AO (LTAO) regime \citep{Oberti:16}. MUSE provides an ideal test case: it is a widely used instrument with a demanding tomographic AO system that is challenging to model, and its broad spectral coverage requires accurate characterization of the polychromatic PSF. These features make MUSE both scientifically important and technically interesting for evaluating the proposed method.
Future papers will investigate scientific applications and extend the framework to other instruments, such as SPHERE \citep{Beuzit:19}.

\subsection{Fourier-based PSF modeling}
As an alternative to PSF-R, Fourier-based PSF modeling was introduced in \citet{Rigaut:98} and later applied in \citet{Jolissaint_synth:10}, \citet{Fusco:20}, \citet{Fetick_PSFAO:19}, and \citet{Neichel:09}. It relies on modeling the power spectral density (PSD) of the AO residuals in the pupil plane, defined in the spatial frequency domain. Instead of reconstructing the AO residuals' PSD from the full telemetry (WFS slopes and DM commands), it computes a theoretical residual-phase PSD from a compact set of integrated atmospheric and system parameters. It includes seeing, coherence time ($\tau_0$), wind profile, guide star positions, and photon flux per WFS subaperture. This makes Fourier-based models far more parsimonious compared to PSF-R, as they rely on a compact set of physically meaningful parameters while retaining broad descriptive power.
Crucially, most of these integrated parameters are routinely stored in the headers of observation FITS files and observatory logs, making them directly accessible from archives and enabling the use of vast amounts of existing data for validation and calibration. Calibration itself is also greatly simplified for Fourier models: a compact parameter space allows these models to be easily optimized to match on-sky data.
In addition, the computational cost is modest, as synthetic residual PSDs can be generated far more efficiently with simple analytical laws.
Given its efficiency, accessibility, and practicality, Fourier-based modeling provides an ideal foundation for the methods developed in this study.

\section{Theory}
\label{sec:theory}
\subsection{PSF formation}
To better understand the input space and the calibration of Fourier-based models, it is useful to first review their theoretical foundation. The focal plane PSF at wavelength $\lambda$ can be computed via the inverse Fourier transform, $\mathcal{F}^{-1}(\,\cdot\,)$, of the optical transfer function (OTF):
\begin{equation}
    I ( \mathbf{x}', \lambda) = F(\lambda) \ \mathcal{F}^{-1}
    \left( \, \mathrm{OTF} \, (\bm{\nu}, \lambda) \, \right) + b(\lambda).
\end{equation}
\noindent
Here,
$\bm{\nu} = (u, v)^\top \in \mathbb{R}^2$ is the spatial frequency vector in OTF space ($u^2+v^2\leq1$), $\mathbf{x}' \in \mathbb{R}^2$ denotes image-plane coordinates,
$F$ is the flux normalization factor, and
$b$ is the constant background offset.
The OTF itself can be factorized into distinct components:
\begin{equation}
    \mathrm{OTF} \, (\bm{\nu}) = \
    D_\mathrm{xy}(\bm{\nu}) \
    \mathrm{OTF}_{\mathrm{static}}(\bm{\nu}) \
    \mathrm{OTF}_{\mathrm{turb}}(\bm{\nu}) \
    \mathrm{OTF}_{\mathrm{TT}}(\bm{\nu}).
    \label{eq:OTF}
\end{equation}
\noindent
For clarity, $\lambda$ is omitted here but will be reintroduced later.
The term $D_\mathrm{xy}$ represents a focal-plane sub-pixel shift, expressed as
\begin{equation}    
    D_\mathrm{xy}(\bm{\nu}) = \exp \left( - 2\pi i \,\frac{D \,\alpha_\mathrm{pix}}{\lambda} \, \left( u \, \delta x + v \, \delta y \right) \right),
\end{equation}
\noindent
where
$\delta x$ and $\delta y$ are sub-pixel displacements in the image plane,
$\alpha_\mathrm{pix}$ is the angular pixel scale on-sky, and
$D$ is the telescope aperture diameter. Adjusting this term enables accurate modeling of astrometric shifts.

The term $\mathrm{OTF}_\mathrm{static}$ in Eq.~(\ref{eq:OTF}) accounts for all quasi-static aberrations in the system and is defined as
\begin{equation}
    \mathrm{OTF}_\mathrm{static} (\bm{\nu}) = \mathcal{P}(\mathbf{x}) \circledast \mathcal{P}^*(-\mathbf{x}), \;\;
    \mathcal{P}(\mathbf{x}) := P(\mathbf{x}) \exp \, \bigl(-i \ \Phi(\mathbf{x}) \bigr),
    \label{eq:OTF_static}
\end{equation}
\noindent
where $\circledast$ denotes the autocorrelation,
$\mathbf{x} \in \mathbb{R}^2$ are pupil-plane coordinates,
$P$ is the pupil mask, and
$\Phi$ is the pupil-plane phase, which can represent NCPAs or other quasi-static aberrations.
This term is revisited in Sect.~\ref{sec:phase_calib}. The rationale for treating $\mathrm{OTF}_\mathrm{static}$ as a separate contributor is outlined in \citet[see Eqs. 1--5]{Jolissaint_synth:10}.

In contrast, $\mathrm{OTF}_\mathrm{turb}$ represents the impact of AO residuals on the infinite-exposure PSF. As is shown in Appendix~\ref{ap:sec:OTF_turb}, it was derived from the PSD of the residual phase $W$, defined in spatial-frequency space $\mathbf{k} = (k_x, k_y)^\top \in \mathbb{R}^2$. Modeling $W$ is therefore central to the Fourier-based framework and is examined in detail in Sect.~\ref{sec:PSD_errors}.

Finally, the last term in Eq.~(\ref{eq:OTF}) accounts for the residual tip–tilt (TT) jitter, which is the random, rapid displacement of the PSF across the focal plane. A 2D Gaussian convolution kernel can approximate its long-exposure effect:
\begin{equation}
    \mathrm{OTF}_\mathrm{TT}(\bm{\nu}) = \exp\left( -\frac{\pi^2}{4 \ln2} \left(\frac{D}{\lambda}\right)^2 \left( J_x^2\,u_r^{\,2} + J_y^2\,v_r^{\,2} \right) \right),
\end{equation}
\[
    \bm{\nu}_r := (u_r, v_r)^\top = \mathbf{T}(\theta_J) \, (u, v)^\top.
\]
\noindent
Here, $J_x$ and $J_y$ denote the FWHM of the tip and tilt jitter, respectively. Unlike the formulation in Eq.~(10) of \citet{BeltramoMartin:20}, the kernel is expressed in a rotated, axis-aligned coordinate frame rather than through the TT covariance matrix, with $\theta_J$ defining the rotation angle and $\mathbf{T}$ the corresponding rotation matrix (see Eq.~(\ref{ap:eq:jitter_rotation})).

Although both $\mathrm{OTF}_\mathrm{TT}$ and $\mathrm{OTF}_\mathrm{turb}$ arise from imperfect AO correction, separating them is well justified. First, in many instruments (including MUSE-NFM), high-order correction is enabled by laser guide stars (LGSs), which are insensitive to TT. Tip and tilt are instead measured by dedicated low-order (LO) WFSs, whose contribution must be modeled separately. Second, Fourier-based models reproduce high spatial frequencies well but poorly capture the low-frequency region of the PSD, which contains information about TT. Finally, isolating the TT component allows for it to be straightforwardly calibrated by tuning $J_x$~and~$J_y$ independently from $\mathrm{OTF}_\mathrm{turb}$.

\subsection{Simulating AO residuals}
\label{sec:PSD_errors}

The key remaining question is how to compute the term $W$ that characterizes the statistics of AO residuals. This derivation begins by identifying the main contributors to imperfect AO correction, as outlined in Sect.~A of \citet{Jolissaint:06} and Sect.~2 of \citet{Correia:17}:
\begin{itemize}
    \item Fitting error: the DM can only correct spatial frequencies, $\mathbf{k}$, below its cutoff frequency $k_c = (2\,d_\mathrm{act})^{-1}$, which is set by the actuator spacing, $d_\mathrm{act}$.
    
    \item WFS noise: arising from photon noise (limited photon flux per integration time) and detector noise.  

    \item Servo-lag error: introduced by the temporal delay between wavefront measurement and DM correction, due to WFS integration and the computation time of the corrective commands.
    
    \item Anisoplanatic error: occurring when the wavefront in the science direction de-correlates from that measured along the guide-star direction with increasing angular separation.  
    
    \item Aliasing error: the WFS samples the wavefront at finite resolution, limited by its Nyquist frequency $k_\mathrm{WFS} = (2\,d)^{-1}$, where $d$ is the subaperture size. Higher spatial frequencies cannot be sensed, and therefore remain uncorrected, folding back into the lower-frequency domain.  
    
    \item Chromatic error: caused when the WFS operates at a wavelength, $\lambda_\mathrm{WFS}$, different from the science wavelength, leading to dispersion-related mismatches.  
\end{itemize}
\noindent
In Fourier-based PSF modeling, these contributions are treated as statistically independent error terms. This framework was first introduced in \citet{Rigaut:98}, extended to two-dimensional and polychromatic formulations in \citet{Jolissaint:06}, \citet{Jolissaint_synth:10}, and \citet{Jolissaint_chromo:10}, adapted to XAO systems in \citet{Correia:14} and \citet{Correia:17}, and further applied in \citet{BeltramoMartin:20}. Its generalization to tomographic systems was developed in \citet{Neichel:09}.

Building on this extensive theoretical groundwork, we provide the explicit expressions of the simulated error terms without derivations. The more detailed explanations are included in Appendix~\ref{ap:sec:PSD_errors}.

\subsubsection{Fitting error}
Since DM cannot correct the spatial frequencies above its cutoff frequency, $k_c$, it is possible to split the residuals PSD into two distinct regions containing corrected ($\left\lVert\mathbf{k}\right\rVert \leq k_c$) and uncorrected ($\left\lVert\mathbf{k}\right\rVert > k_c$) spatial frequencies, where a pure open-loop atmospheric spectrum, $W_\varphi$, governs the latter region (see Eq.~(\ref{ap:eq:von_Karman}). Then, $W_{\mathrm{fitting}}$ can be expressed as
\begin{equation}    
    W_{\mathrm{fitting}}(\mathbf{k}) =
    \begin{cases}
        \; 0 & \mathrm{if } \: \left\lVert\mathbf{k}\right\rVert \leq k_c, \\
        \: \widetilde{\Pi}(\mathbf{k}) \,W_\varphi (\textbf{k}) & \mathrm{otherwise}.
    \end{cases}
    \label{eq:W_fit}
\end{equation}
\noindent
Here, $\widetilde{\Pi}$ is the piston mode filter (see Appendix~\ref{ap:sec:PSD_errors:fitting} for more details).

\subsubsection{Wavefront sensor noise error}
This error term originates from noise in the WFS detectors, which propagates through the wavefront reconstruction. Since MUSE-NFM relies on tomographic correction, the expression here can be adapted from the second term of Eq.~(23) in \citet{Neichel:09}, yielding
\begin{equation}
    W_\eta(\mathbf{k}) = \widetilde{\Pi}(\mathbf{k}) \left(\mathbf{P}_\theta^{\mathrm{DM}} \mathbf{W} \right) \, \mathbf{C}_b \, \left(\mathbf{P}_\theta^{\mathrm{DM}} \mathbf{W} \right)^\top,
\label{eq:W_noise}
\end{equation}
\noindent
where
$\mathbf{P}_\theta^{\mathrm{DM}}$ is the vector \footnote{For on-axis LTAO correction, where a single DM is conjugated to the ground layer, $\mathbf{P}_\theta^{\mathrm{DM}} = \bm{1}$.} that projects the DM phases seen in the direction, $\bm{\theta}$, of the science object in the pupil plane, in which the tomographic solution is optimized (see Eq.~(10) from \citealt{Neichel:09}),
$\mathbf{W}$ is the phase volume reconstruction matrix defined in Eq.~(13) in \citet{Neichel:09}, and
$\mathbf{C}_b$ is the WFS noise variance matrix (see Eq.~(\ref{ap:eq:C_b})).

\subsubsection{Spatio-temporal error}
This error term represents a combination of anisoplanatic\footnote{In tomographic AO, there is no anisoplanatism in the strict SCAO sense. Instead, it is incorporated into the tomographic reconstruction.} and servo-lag errors. As was noted in \citet{Jolissaint_synth:10}, these two contributions are coupled. Under the frozen-flow approximation (Taylor hypothesis), a temporal shift of the wavefront can be partially compensated for by a spatial shift of the opposite sign. The first term of Eq.~(23) in \citet{Neichel:09} can therefore be adapted to describe this error in terms of a PSD:
\begin{equation}
    W_\mathrm{ST\,}(\mathbf{k}) = \widetilde{\Pi}(\mathbf{k}) \: \mathbf{U} \:  \mathbf{C}_\varphi \, \mathbf{U}^\top, \;\;
    \mathbf{U} := \mathbf{P}_\theta^L - \mathbf{P}_\theta^{\mathrm{DM}} \mathbf{W} \mathbf{M} \mathbf{P}_\alpha^L.
    \label{eq:W_STA}
\end{equation}
\noindent
Here, $\mathbf{P}_{\theta}^L$ projects turbulent layers in the direction, $\theta$, onto the pupil plane,
$\mathbf{P}_{\alpha}^L$ projects turbulent layers in the directions, $\alpha$, of the guide stars (GSs),
$\mathbf{M}$ is the WFS measurement operator matrix, originally defined in \citet{Rigaut:98} and \citet{Jolissaint:06} and later extended to tomographic systems in Sect.~6 of \citet{Neichel_thesis:09}.
Further details on these quantities are provided in Appendix~\ref{ap:sec:PSD_errors:spatio_temporal}.
Note that $\mathbf{P}_\theta^L$, $\mathbf{P}_\alpha^L$, $\mathbf{P}_\theta^{\mathrm{DM}}$, $\mathbf{M}$, $\mathbf{W}$, and $\mathbf{C}_\varphi$ are all functions of $\mathbf{k}$, though the notation is omitted here for brevity.

\subsubsection{Aliasing}
\label{sec:theory:aliasing}
As was noted earlier, this error arises when the WFS under-samples the high-spatial-frequency component of the wavefront. Because the wavefront is sampled by a finite number of subapertures or pixels, spatial frequencies above the WFS Nyquist limit are not measured and instead fold back into lower frequencies during discretization. The control system then attempts to correct these spurious low-frequency signals, introducing an aliasing error. 

For this term, we departed from the tomographic framework and adopted an SCAO formulation. Although aliasing derivations for tomographic reconstruction are presented in \cite[Sect. 7.6.4 and Annexe B.1]{Neichel_thesis:09}, the SCAO treatment is computationally more efficient.

Accordingly, we adopted Eq.~(19) from \citet{Correia:17}, re-expressing its terms with the notation of \citet{Correia:14}, to obtain
\begin{equation}
    W_\mathrm{A}(\mathbf{k}) =  \sum\limits_{m} \sum\limits_{n} \left|Q(\mathbf{k}) \ A_\varphi (\mathbf{k}) \right|^2 \, \widetilde{\Pi}(\mathbf{k} + \mathbf{s}/d) \, W_\varphi(\mathbf{k} + \mathbf{s}/d),
\end{equation}
\noindent
where $\mathbf{s} = (n, m)^\top \in \mathbb{N}^2 \setminus \{\,(0,0)^\top\}$; $Q$ and $A_\varphi$ functions are defined in Appendix~\ref{ap:sec:PSD_errors:aliasing}.

\subsubsection{Chromatism error}
The PSD terms described above are assumed to be wavelength-independent (neglecting atmospheric dispersion) for computational simplicity. In practice, however, weak dispersion is always present and must be considered, particularly for strongly multispectral instruments such as MUSE.

One such effect is the chromatism error. Following Eqs.~(12)–(13) of \citet{Fusco:06} and Eq.~(3) of \citet{Jolissaint_chromo:10}, it can be written as
\begin{equation}
    W_{\Delta n}(\mathbf{k}, \lambda) = \left(1 - \frac{\Delta n(\lambda)}{\Delta n(\lambda_{\mathrm{GS}})}\right)^2  \widetilde{\Pi}(\mathbf{k}) \, W_\varphi(\mathbf{k}).
\end{equation}
\noindent
Here,
$\lambda_{GS}$ is the WFS wavelength,
$\Delta n(\lambda) := n(\lambda)- 1$ is the refractivity, and
$n(\lambda)$ is the refractive index of air according to \citet{Ciddor:96}.
See Appendix~\ref{ap:chroma} for a more detailed description.

\subsubsection{Differential refraction error}
A second chromatic error contributor is the differential refraction error. This effect can be formulated as an anisoplanatism-like term by simplifying Eqs.~(7)–(9) of \citet{Jolissaint_chromo:10}. As with $W_\mathrm{A}$, we adopted an SCAO formulation for computational efficiency, yielding
{\setlength{\jot}{3pt}
\begin{equation}
    \begin{aligned}
        W_\mathrm{DR}(\mathbf{k}, \lambda) =
        \sum_{l=1}^{N_L}
        2\,\mathbf{w}[l]\,
        \bigl(1-\cos\bigl(\mathbf{k}\cdot\mathbf{\Theta}_{\mathrm{DR},\,l\,}(\lambda)\bigr)\bigr)\,
        \widetilde{\Pi}(\mathbf{k})\,W_\varphi(\mathbf{k})
    \end{aligned}
    \label{eq:diff_ref}
.\end{equation}}\unskip
\noindent
For more details, see Appendix~\ref{ap:diff_ref}.

\subsubsection{Computing the total AO residuals' PSD}
Assuming the statistical independence of the individual PSD terms described above, the total PSD $W_\mathrm{total}$ was computed as the sum of the individual contributors:
{\setlength{\jot}{8pt}
\begin{equation}
    \begin{aligned}
        W(\lambda) &= \left( \frac{\lambda_\mathrm{atm}}{2 \pi} \right)^2
        \Big( W_{c}(\lambda) \Big|_{\left\lVert\mathbf{k}\right\rVert \, \leq \, k_c}
             + W_\mathrm{fitting} \Big|_{\left\lVert\mathbf{k}\right\rVert \,>\, k_c} \, \Big), \\
        W_{c}(\lambda) &:= W_\mathrm{A} + W_\mathrm{ST} + W_\eta
            + W_{\Delta n}(\lambda) + W_\mathrm{DR}(\lambda) + \Delta W.
    \end{aligned}
\label{eq:PSD_total}
\end{equation}}\unskip
\noindent
In Eq.~(\ref{eq:PSD_total}), $\mathbf{k}$ is omitted for clarity. The prefactor $(\lambda_\mathrm{atm} / 2 \pi)^2$ converts the residual phase PSD from $\mathrm{rad}^2\,\mathrm{m}^{-2}$ into an optical path difference (OPD) PSD in square nanometers per square meter. The additional term, $\Delta W$, represents unexplained residual variance. Its role is discussed later, in Sect.~\ref{sec:calib_PSD}.
\begin{figure}[h!]
    \centering
    \includegraphics[width=0.485\textwidth]{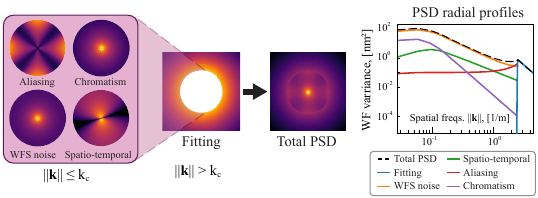}
    \caption{
        Illustration of Eq.~(\ref{eq:PSD_total}). $W_\mathrm{DR}$ and $\Delta W$ are omitted for brevity.
    }
    \label{fig:PSD_scheme}
\end{figure}

With this, the theoretical framework is complete. The identification of the main contributors to AO residuals, together with their corresponding analytical expressions, now provides the basis for defining the input space of the PSF model.

\subsection{Identifying model inputs}
\label{sec:model_inputs}
Based on the theoretical framework outlined in Sect.~\ref{sec:theory} and Appendix~\ref{ap:sec:theory}, we identified the set of variable dynamic parameters, $\mathcal{Y}_d$, that define the PSF under different observing conditions for a given instrument and AO regime. These parameters vary from one observation to another:
\begin{equation}
    \mathcal{Y}_d = 
    \left\{\
    \begin{aligned}
         &J_x(\lambda),\;J_y(\lambda),\;\delta x(\lambda),\;\delta y(\lambda),\; F(\lambda),\;b(\lambda),\\
         &\Phi(\mathbf{x},\lambda),\;\theta_J,\;\theta_{\mathrm{zen}},\;\bm{\theta},\;\sigma^2_{\mathrm{ph}},\;\Delta\sigma^2,\\
         &r_0(\lambda),\;L_0,\;N_L,\;\mathbf{h},\;\mathbf{w},\;\bm V
    \end{aligned}
    \ \right\}.
\end{equation}
\noindent
A second set, $\mathcal{Y}_s$, contains static parameters that are intrinsic to MUSE-NFM and remain fixed across observations. These are well-characterized system properties:
\begin{equation}
    \mathcal{Y}_s = 
    \left\{\
    \begin{aligned}
         &d_{\mathrm{act}},\;d,\;\lambda_{\mathrm{atm}},\;\lambda_{\mathrm{GS}},\;N_{\mathrm{GS}},\;\bm{\alpha},\;P(\mathbf{x}),\\
         &\;\alpha_{\mathrm{pix}},\;\sigma^2_{\mathrm{det}},\;D,\;\Delta t,\;\zeta
    \end{aligned}
    \ \right\}.
\end{equation}
\noindent
Assuming $\Phi = \mathbf{0}$ and $\Delta W = \mathbf{0}$, the union $\mathcal{Y}_d \cup \mathcal{Y}_s$ fully determines the PSF. Notably, almost all elements of these sets are integrated quantities that can be derived from raw AO telemetry. In practice, model calibration and tuning to specific atmospheric conditions require adjusting only $\mathcal{Y}_d$ given the same instrument and AO regime. The glossary with all parameter definitions can be found in Table~\ref{tab:glossary_variables}.

This discussion further highlights both the data efficiency and the flexibility of the Fourier modeling approach. Moreover, clarifying the input space establishes the basis for the calibration methodology presented in the following section.

\section{Data-calibrated PSF modeling}
\subsection{Motivation}

The ease of calibration was previously identified as one of the key factors in selecting the Fourier-based PSF modeling framework. Hence, it is essential to clarify the notion of calibration in this context and motivate it.
\begin{itemize}
    \item Uncertainty in input parameters: In practice, integrated parameters are often imperfectly known and affected by measurement errors (see Fig.~8 of \citealt{Fetick_PSFAO:19}).
    \item Telemetry processing: Since these parameters are derived from raw telemetry, this process can introduce errors.
    \item Model simplifications: The Fourier-based approach necessarily relies on approximations, whereas real AO systems are highly complex and include many poorly characterized, sometimes dynamic, effects. Calibration allows the PSF model to absorb these discrepancies from data to provide a more realistic PSF representation.
\end{itemize}
To summarize, in this context, calibration means compensating for systematic biases and unmodeled effects in the data and model to achieve a more realistic representation of PSF morphology. Its role is further demonstrated in Appendix~\ref{ap:calib_tuned_direct}. Here, we first specify which components of the model are subject to calibration.

\subsection{Identifying calibratable quantities}
From the analysis in Sect.~\ref{sec:theory}, four domains emerge as candidates for calibration:
\begin{itemize}
    \item the vector of input variables, $\mathcal{Y}_d$,
    \item the static pupil-plane phase, $\Phi(\mathbf{x},\lambda)$,
    \item the AO residuals' PSD, $W(\mathbf{k}, \lambda)$,
    \item the focal-plane intensity distribution, $I(\mathbf{x}')$.
\end{itemize}
In this case, adjusting $\mathcal{Y}_d$ propagates through all four domains. In addition, a phase term, $\Phi$, can capture NCPAs, and an additive PSD term, $\Delta W$, can represent unexplained residual variance. In principle, a focal-plane correction, $\Delta I$, could account for detector-specific artifacts, but here we assume that no such artifacts are present.

\subsubsection{Calibrating pupil phase}
\label{sec:phase_calib}
\begin{figure}[h!]
    \centering
    \includegraphics[width=0.5\textwidth]{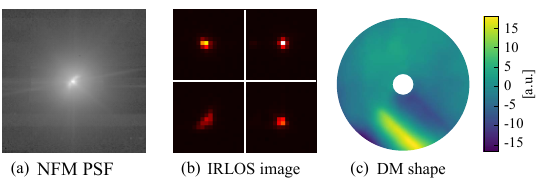}
    \caption{
        (a) Effect of the phase bump on MUSE-NFM PSFs (log scale).
        (b) Phase bump observed by the IRLOS WFS.
        (c) Measured phase bump.
    }
    \label{fig:MUSE_phasebump}
\end{figure}

In the case of MUSE-NFM, static phase calibration must account for a specific instrument-related effect known as the “phase bump.” This NCPA appeared between late 2021 and early 2024 due to incorrect reference-slope calibration in one of the WFSs, caused by background biases in a specific octant of the LGS WFS detector. It manifested as a quasi-static phase structure (Fig.~\ref{fig:MUSE_phasebump}) whose strength varied from observation to observation.
\noindent
The morphology of the phase bump was measured separately (see Fig.~\ref{fig:MUSE_phasebump}c) and incorporated in the PSF model in the pupil phase. Then, the term $\Phi$ in Eq.~(\ref{eq:OTF_static}) can now be defined as
\begin{equation*}
    \Phi(\mathbf{x}, \lambda) = \frac{2 \pi}{\lambda} \, P_b(\mathbf{x}) \ a_b,
\end{equation*}
\noindent
where $P_b$ is the pupil-plane OPD map of the measured phase bump (Fig.~\ref{fig:MUSE_phasebump}c), and $a_b$ is a scaling factor that sets its amplitude. Other NCPAs may also contribute to the pupil phase, as is discussed in \citet{Kuznetsov:24}. For simplicity, however, we assume here that the phase bump is the sole source of quasi-static phase error.

\subsubsection{Calibrating residuals PSD}
\label{sec:calib_PSD}
To account for errors in the residual PSD, we adopted a simplified 2D Moffat function \citep{Moffat:69} as a flexible and calibratable ad hoc PSD error absorber. It can be expressed as
\begin{equation}
    \Delta W(\mathbf{k}) =
    \widetilde{\Pi}(\mathbf{k}) \left( \frac{A \, (\beta_M - 1) \,  \alpha_M^{2(\beta_M-1)} }{\pi \left(\alpha_M^2 + \left\lVert\mathbf{k}\right\rVert^2 \right)^{\beta_M}} + B \, \right),
\end{equation}
\noindent
where $A$ and $B$ denote amplitude and background, while $\alpha_M$ and $\beta_M$ control the function's width and steepness, respectively.

\subsubsection{Calibrating model inputs}
\label{sec:inputs_calib}
Absorbers act as subtle, ad hoc extensions to the otherwise fully physics-based PSF model, enhancing its representativeness. However, like the primary model inputs, they must also be calibrated. Because both $\Phi$ and $\Delta W$ are expressed in parametrized form, their corresponding parameters can be incorporated into the model’s input space. Thus, the entire problem of PSF model calibration reduces to calibrating the model's input domain. We therefore defined the set of calibratable parameters, $\mathcal{Y}_c$, as
\[
    \mathcal{Y}_c = \{ \,
        F(\lambda), \,
        J_x(\lambda), \,
        J_y(\lambda), \,
        \Delta \sigma^2, \,
        r_0, \,
        \mathbf{w},
        A, \,
        B, \,
        \alpha_M, \, 
        a_b \,
    \}.
\]
\noindent
This set includes only a subset of inputs from $\mathcal{Y}_d$. For instance, $\beta_M$ and $\theta_J$ are not included, as they are degenerate with other inputs and cannot be uniquely recovered. Likewise, $\bm{\theta}$, $\delta x(\lambda)$, $\delta y(\lambda)$, and $b(\lambda)$ are not calibratable, since they vary unsystematically for different targets. The systematic shifts in $\delta x,\delta y$ due to dispersion and field distortion are effectively corrected by the MUSE data reduction pipeline. The error in the determination of photon noise variance, $\sigma^2_{\mathrm{ph}}$, is absorbed by the $\Delta \sigma^2$ term. Telescope pointing is assumed to be perfectly known. Varying the number of turbulence layers, $N_L$, breaks the model's differentiability, and optimizing the height of the atmospheric layers, $\mathbf{h}$, introduces strong coupling with $\mathbf{w}$. Finally, we excluded $\mathbf{V}$ and $L_0$, whose effect on the PSF is minor in the case of MUSE-NFM and could lead to overfitting.

\subsection{Calibration dataset}
\label{sec:calib_dataset}
After defining the model input space and calibratable variables, it is necessary to identify the sources of parameter values. As was mentioned previously, some quantities can be found in the reduced telemetry archive, which encompasses the following data sources:
\begin{itemize}
    \item FITS headers of observations (after running ESO's MUSE data reduction pipeline). This is the primary source of information on observation conditions and parameters. 
    \item Raw FITS files of observations before running the pipeline. They contain additional information that is lost during the data reduction pipeline. For example, raw headers contain $C_n^2$ profiles or LO WFS images.
    \item Logs generated by the UT4 telescope and the MUSE instrument. They are used to complement information missing from FITS headers, such as LO AO loop parameters, natural guide star (NGS) parameters, and other relevant entries.
\end{itemize}
\noindent
To implement the calibration, a set of real on-sky PSFs is required as a reference. These PSFs can then be paired with reduced telemetry to form the calibration dataset. For this purpose, we used spectrophotometric standard stars, which have been routinely observed with MUSE-NFM since 2018 and are archived in the ESO science database. These observations capture bright stars under a wide range of atmospheric conditions. Their abundance, high signal-to-noise ratio (S/N), and strictly point-source nature make them an excellent reference for PSF model calibration. The resulting dataset, $\mathcal{D}$, was then defined as
\[
    \mathcal{D} = \big\{ \, \bigl(\mathcal{X}^{(k)}, \ I_\mathrm{data}^{(k)}(\mathbf{p}, \lambda) \bigr) \, \big\}^{N_I}_{k=1}.
\]
\noindent
Each $k$-th sample in the dataset consists of an observed PSF, $I_\mathrm{data}$, and a corresponding set of reduced telemetry, $\mathcal{X}$, recorded during the exposure. In total, $N_I = 392$ samples were collected, excluding those with corrupted PSFs or missing telemetry. The continuous image-space coordinates, $\mathbf{x}' \in \mathbb{R}^2$, are represented here by discrete focal-plane pixel indices, $\mathbf{p} \in \Omega$, where $\Omega={1,\dots,H}\times{1,\dots,W}$ denotes the pixel grid of an image with a height, $H$, and width, $W$. All $I_\mathrm{data}$ images were background-subtracted and spectrally binned. The original $N_\lambda = 3681$ chromatic slices from MUSE-NFM were averaged into $N_\lambda = 7$ evenly spaced bins of $\Delta \lambda \approx 65$ nm, improving the S/N and reducing computational cost.

Here, we also introduced an additional set of parameters, $\mathcal{X}$. In contrast to $\mathcal{Y}_s$ and $\mathcal{Y}_d$, which define the PSF model input space, $\mathcal{X}$ comprises integrated telemetry recorded by various instrument subsystems ($\mathcal{X} \neq \mathcal{Y}_d \, \cup \, \mathcal{Y}_s$). These include
\begin{itemize}
    \item atmospheric parameters ($r_0$, $L_0$, $\tau_0$, full $C_n^2$ profile, $\mathbf{V}$ at different altitudes), which are measured both by the AO system of UT4 and by DIMM/MASS-DIMM instruments;
    \item LO and LGS loop parameters (loop gain and frequency, guide star fluxes);
    \item telescope pointing ($\theta_\mathrm{zen}$, airmass);
    \item ambient conditions (pressure, temperature, humidity);
    \item other auxiliary quantities (such as the RMS of WFS slopes and more).
\end{itemize}
At the same time, $\mathcal{X}$ excludes some model-specific and non-systematic quantities such as $\delta x$, $\delta y$, and $b$, as well as parameters that are calibrated ($\mathcal{X} \cap \mathcal{Y}_c = \varnothing $). As a result, $\mathcal{X}$ is redundant in the sense that it contains more logged parameters than is strictly required by the PSF model, yet it is incomplete with respect to the full set of model inputs.

\subsection{Calibrated PSF prediction}
The structure of the problem, with its data redundancy and the use of ad hoc error absorbers, naturally implies a data-driven calibration strategy based on machine learning (ML). In this framework, the calibration of the PSF model can be formulated as the training of a calibrator function that predicts\footnote{Throughout this work, "prediction" does not denote forecasting, but simply the inference by a trained calibrator model.} the values of calibrated inputs, $\mathcal{Y}_c$, that minimize the discrepancy between the observed data $I_\mathrm{data}$ and the model. We further hypothesize that the redundant entries in $\mathcal{X}$ may contain latent information that enables the calibrator to capture relationships in the data that are not analytically tractable.

The resulting calibrated PSF model prediction can thus be expressed as
\begin{equation}
        \widetilde{I} (\mathbf{x}', \lambda) = f(\mathbf{x}', \lambda; \, \mathcal{Y}_s \cup \mathcal{Y}_d \oplus \widetilde{\mathcal{Y}_c}),
        \quad
        \widetilde{\mathcal{Y}}_c = g\,(\mathcal{X};\, \gamma).
        \label{eq:PSF_model_f}
\end{equation}
\noindent
Here,
$f(\,\cdot\,)$ is the PSF model,
$g(\,\cdot\,)$ is the calibrator with trainable parameters, $\gamma$,
the “$\,\widetilde{\,\cdot\,}\,$” notation stands for predicted quantities,
$\widetilde{\mathcal{Y}}_c$ is the predicted set of calibrated model inputs,
$\widetilde{I} (\mathbf{x}', \lambda)$ is the predicted PSF, and
$\oplus$ is the right-biased union operator, which updates overlapping entries from $\mathcal{Y}_d$ with the values from $\mathcal{Y}_c$.
For two arbitrary sets, $A$ and $B$, it is defined as $A \oplus B = \{\, A \setminus B \, \} \, \cup \, B$.
Note that the values in $\mathcal{X}$ are standardized (zero mean and unit variance) before being input to the calibrator.

\subsection{Implementation}
\label{sec:PSF_pred:implementation}
\begin{figure}[ht!]
    \centering
    \includegraphics[width=0.49\textwidth]{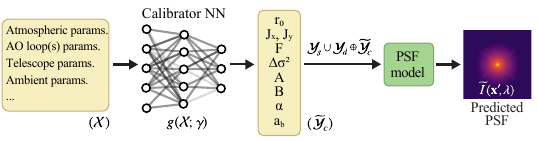}
    \caption{Schematic of the data-calibrated PSF predictor for MUSE-NFM.}
    \label{fig:MUSE_calibrator}
\end{figure}

After outlining the general concept behind the calibration approach, the actual implementation can be discussed. In this work, we propose a hybrid approach introduced in \citet[see our Fig.~\ref{fig:MUSE_calibrator}]{Kuznetsov:23} that combines a data-driven calibrator, implemented as a neural network (NN), followed by a fully analytical PSF model that implements the principles outlined earlier in Sect.~\ref{sec:theory}.

This hybrid design is motivated by the limited availability of independent on-sky training data. Although $\mathcal{X}$ spans a high-dimensional parameter space, the size of the training set, $\mathcal{D}$, is small, necessitating strong regularization during calibrator training. This is achieved by limiting the calibrator NN to learning only the low-dimensional mappings between $\mathcal{X}$ and $\widetilde{\mathcal{Y}_c}$, while the analytical model accounts for the physics of PSF formation. It keeps the number of trainable parameters small. In addition, because the calibrator is directly coupled to a physics-driven model, only physically meaningful predictions result in realistic PSFs, which provides an additional source of regularization.

As a bonus, in this approach, the calibrator can be analyzed independently of the PSF model using feature-attribution tools, such as SHAP \citep{Lundberg:17}, to probe the learned interactions among the reduced telemetry and model inputs. This aspect will be further explored in future work.

An NN-based solution was chosen for the calibrator due to its flexibility and ability to approximate smooth, differentiable functions efficiently\footnote{The proposed scheme is not limited to NNs and can utilize other types of regression models, provided that they are differentiable.}.
The architecture is a feed-forward NN with three hidden layers of 200 neurons each, using hyperbolic tangent activation functions. The input and output dimensionalities are 61 and 27, respectively, and a dropout layer with a probability of 10\% is applied before the final layer.

An obvious limitation of the proposed implementation is its restricted ability to reproduce intricate, unexplained morphological features of the PSF due to the strong parametrization assumptions imposed on the $\Delta W$ and $\Phi$ terms. However, such structures were not observed on MUSE-NFM PSFs. Moreover, as more training data accumulates in the future, additional optimizable degrees of freedom could be introduced, enabling more flexible error-absorbing components.

In the long term, a fully data-driven approach could be pursued. For example, by employing a generative model (as in \citealt{Jia:21} and \citealt{Long:20}) to infer the PSF directly from reduced telemetry. Such a model, however, would require orders of magnitude more training data than is currently available. One possible pathway would be to pretrain such a model on synthetic data from end-to-end AO simulations and subsequently fine-tune it on on-sky observations. While this strategy would maximize flexibility by learning unexplained effects directly from data without relying on strong analytical assumptions, it would do so at the cost of interpretability and significantly higher computational demands during training.

\subsubsection{Training the calibrator}
The data-driven calibrator, $g(\mathcal{X}^{(k)}; \gamma)$, was trained to minimize the loss function, $\mathcal{L}$, 
between the measured PSFs and the corresponding model predictions, where $\gamma$ denotes the trainable weights and biases of the calibrator. In total, 314 samples from $\mathcal{D}$ were used for training, while 78 were reserved for validation. Further details are provided in Appendix~\ref{ap:calib_train}.

\begin{figure}[ht!]
    \centering
    \includegraphics[width=0.5\textwidth]{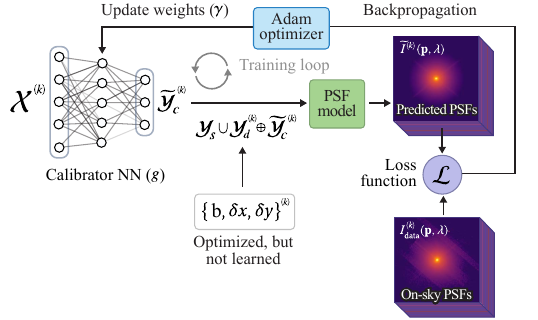}
    \caption{Training data-driven PSF predictor.}
    \label{fig:calib_training}
\end{figure}

During the training, as is depicted in Fig.~\ref{fig:calib_training}, parameters ${b(\lambda), \delta x, \delta y}$ are optimized but are not predicted by the calibrator, since they do not vary systematically (as is briefly mentioned in Sect.~\ref{sec:calib_dataset}). The calibrator is optimized jointly with the analytical PSF model: gradients are backpropagated through the physics-based model, which acts as a regularizer.

\subsubsection{Code}
Implementing the proposed hybrid approach, which combines ML with analytical modeling, required the development of dedicated software. To this end, we developed TipTorch, a novel open-source simulation tool (see the software availability section at the end of the paper).

The analytical core of TipTorch is inspired by the P3 library, now integrated into the astro-TIPTOP package \citep{Neichel:21}, which currently serves as the baseline PSF simulation tool for the ERIS, MAVIS, and HARMONI instruments. However, one of TipTorch's key features is its full differentiability. This is achieved with automatic differentiation (AD), implemented by leveraging the PyTorch framework \citep{Paszke:19}. A similar idea of differentiable PSF modeling combined with learned corrections was also previously explored in \citet{Liaudat:23}, \citet{Desdoigts:23}, and \citet{Desdoigts:24}. In our case, using AD is essential for enforcing physics-based regularization via backpropagation through the PSF model.

In addition to AD, the PyTorch-based implementation supports highly parallelized simulations that can be flawlessly executed on both the CPU and GPU backends, enabling TipTorch to simulate dozens of polychromatic PSFs per second on desktop-level hardware\footnote{The NVIDIA RTX A4000 Laptop GPU was used for these tests.}. The code's efficiency enables coverage of the full MUSE-NFM wavelength range and the simulation of dense stellar fields.

\section{Results}

\subsection{Standard stars}
\label{sec:results:STD}
We assessed the performance of the proposed method on the validation subset of $\mathcal{D}$.
Table~\ref{tab:calib_error_metrics} summarizes how well the calibrated model reproduces the on-sky PSF peak, core width, and overall PSF structure.

\begin{figure}[h!]
    \centering
    \includegraphics[width=0.485\textwidth]{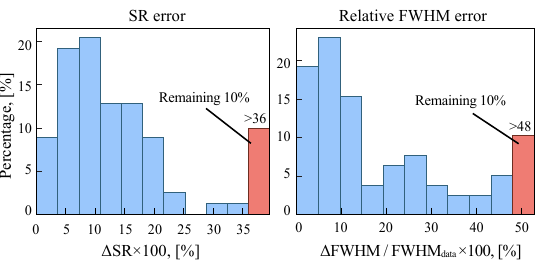}
    \caption{
        Distributions of SR (left plot) and FWHM (right plot) errors for the calibrated model.
    }
    \label{fig:MUSE_calibrated_SR_FWHM}
\end{figure}

\begin{table}[h!]
    \centering
    \renewcommand{\arraystretch}{1.2}
    \setlength{\tabcolsep}{10pt}
    \caption{Median validation metrics for the calibrated model. Lower values indicate better performance.}
    \label{tab:calib_error_metrics}
    \begin{tabular}{lc}
        \hline\hline
        Metric & Median value \\
        \hline
        $\Delta\mathrm{SR}$, [\%] & 13.5 \\
        $\Delta \mathrm{FWHM}_{\mathrm{rel}}$, [\%] & 10.9 \\
        $\Delta \mathrm{FWHM}$, [mas] & 9.1 \\
        FVU, [\%] & 3.2 \\ \hline
    \end{tabular}
\end{table}
\noindent
Here, the Strehl ratio error ($\Delta$SR) denotes the peak difference between predicted and on-sky PSFs, while the fraction of variance unexplained (FVU) was computed as in Eq.~(\ref{ap:eq:FVU}). The relative and absolute FWHM errors, $\Delta \mathrm{FWHM}_\mathrm{rel}$ and $\Delta \mathrm{FWHM}$, respectively, quantify the difference in PSF core width; the FWHM was derived from a 1D Moffat profile fit to radial cuts of $I_\mathrm{data}$ and $\widetilde{I}$ within 5 pixels of the PSF center. For each validation sample, metrics were first calculated per (binned) spectral slice, then averaged, and finally computed across the dataset by taking the median. More information on SR and FWHM error distributions can be found in Fig.~\ref{fig:MUSE_calibrated_SR_FWHM}.

\begin{figure}[h!]
    \centering
    \includegraphics[width=0.485\textwidth]{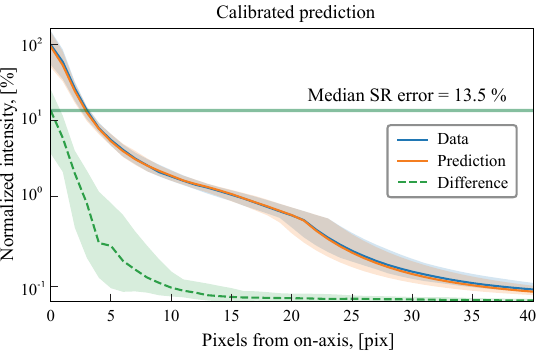}
    \caption{
        Median radial PSF profiles of the validation dataset. Error bars indicate the 1$\sigma$ dispersion across the dataset. The profiles are averaged across all wavelengths. The apparent “kink” in the curves arises from the logarithmic scaling of the plot, which transitions to a linear scale just below the 1\% level.
    }
    \label{fig:MUSE_profiles_calibrated}
\end{figure}

Before metrics evaluation, both observed and predicted PSFs were normalized per spectral slice to the unit flux,
\begin{equation*}
    \sum_{\mathbf{p} \, \in \, \Omega} I_\mathrm{data} (\mathbf{p}, \lambda) = 1, \;
    \sum_{\mathbf{p} \, \in \, \Omega} \widetilde{I} (\mathbf{p}, \lambda) = 1.
\end{equation*}
For illustration, Fig.~\ref{fig:MUSE_profiles_calibrated} shows the median radial PSF profiles from the validation set, computed with the \texttt{photutils} Python library. Profiles were normalized to the peak of the observed PSF and scaled to percent.
Both Table~\ref{tab:calib_error_metrics} and Fig.~\ref{fig:MUSE_profiles_calibrated} demonstrate an agreement between the predicted and on-sky PSFs, indicating a good overall performance of the method (see also Appendix~\ref{ap:calib_tuned_direct} for a comparison with the uncalibrated case).

\begin{figure}[h!]
    \centering
    \includegraphics[width=0.5\textwidth]{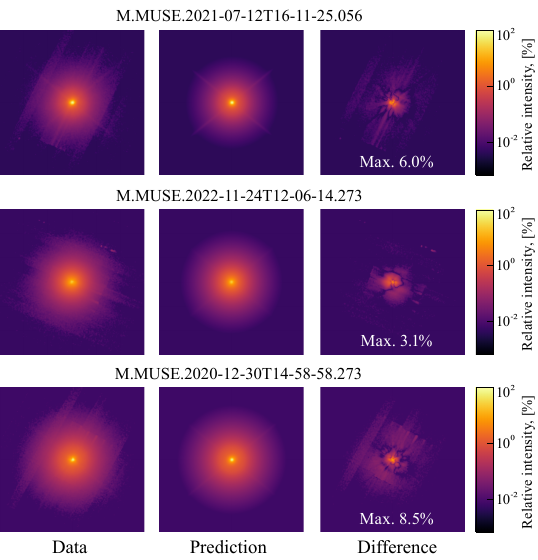}
    \caption{
        Three example PSFs from the validation dataset. Left: Data. Middle: Calibrated prediction. Right: Residuals. All images are shown on a logarithmic scale. All PSFs are spectrally averaged.
    }
    \label{fig:MUSE_calibrated_examples}
\end{figure}

\subsection{$\omega$ Centauri cluster}
\begin{figure}[h!]
    \centering
    \includegraphics[width=0.485\textwidth]{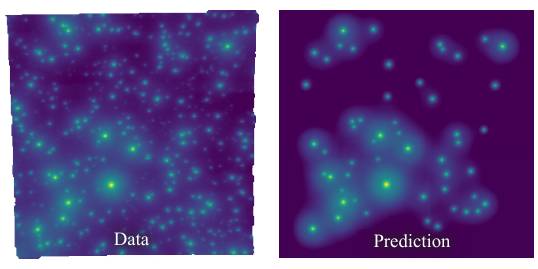}
    \caption{
        Comparison between observed data (left) and predicted PSFs (right) for $\omega$ Centauri.
    }
    \label{fig:MUSE_omega_1}
\end{figure}

\begin{figure}[h!]
    \centering
    \includegraphics[width=0.485\textwidth]{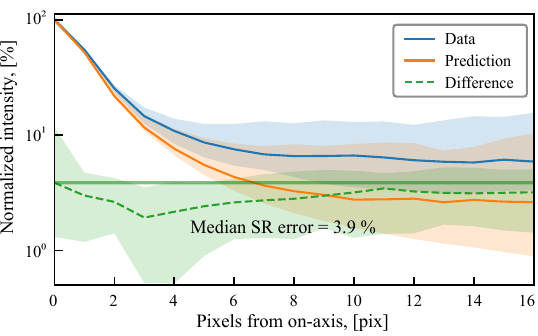}
    \caption{
        Normalized radial PSF profiles for the $\omega$ Centauri field. The profiles were computed in the same way as in Fig.~\ref{fig:MUSE_profiles_calibrated}. Only astrometric shifts $\delta x, \delta y$ were fit; no other PSF parameters were adjusted.
    }
    \label{fig:MUSE_omega_2}
\end{figure}

The validation dataset described above is limited in scope and serves mainly as a proof of concept. To further assess the method on scientifically relevant data, we applied it to observations of the crowded stellar field in the core of the $\omega$ Centauri cluster. The dataset, provided by Sebastian Kamann \citep{Pechetti:24}, is particularly well suited for testing because a single exposure contains multiple point sources, enabling an assessment of both prediction accuracy and field-dependent variability within the $7.5\arcsec \times 7.5\arcsec$ field of view. Figure~\ref{fig:MUSE_omega_1} shows the PSF prediction results for the 54 brightest sources in the field.
Radial profiles, computed in the same way as for Fig.~\ref{fig:MUSE_profiles_calibrated}, are shown in Fig.~\ref{fig:MUSE_omega_2}.
The results demonstrate an excellent predictive performance, with a median $\Delta$SR of less than 5\%. Accuracy is higher than for the standard-star dataset, likely because the observing conditions in this dataset are well represented in the training data, whereas the standard-star validation set covers a broader range of conditions. The observation was obtained at a seeing of $0.54\arcsec$, with an NGS J-band magnitude of 13.6 and an isoplanatic angle of $1.9\arcsec$.
In Fig.~\ref{fig:MUSE_omega_2}, residual “flat” discrepancies are primarily due to the faint background sources that were not included in the simulation. 
Although the calibrator was not explicitly trained to capture off-axis behavior, the observed off-axis PSF variability is fully reproduced by the analytical component of the model.

\section{Discussion}
Although the present study is optimized for MUSE-NFM, the approach remains general. As will be shown in forthcoming papers in the series, the analytical PSF model can simulate a wide range of correction AO regimes. In this case, instrument specificity is attributed to the data-driven calibrator, as the set of reduced-telemetry inputs is unique to each instrument. However, given sufficient data, the same training logic can be applied to any system. Forthcoming work will demonstrate the same approach applied to SPHERE (the first results were already reported in \citealt{Kuznetsov:23}).

Once trained, the calibrator is stable for operational use, but it can still be fine-tuned as new data becomes available. For MUSE-NFM, current training is biased toward bright on-axis stars due to the use of the standard spectro-photometric stars dataset. Thus, extending the training to faint and off-axis sources is a natural next step to explore in forthcoming papers.
Therefore, regularly acquiring moderately crowded stellar fields, similar to Fig.~\ref{fig:MUSE_omega_1}, would usefully complement the current datasets to facilitate off-axis calibration. Beyond the needs of this work, such data would also aid in debugging and calibrating future PSF–modeling approaches, including ones based on classical PSF-R.

Despite the message at the beginning of the paper, Fourier modeling and PSF-R should be viewed as complementary rather than competing approaches. Fourier models efficiently explore parameter space using readily accessible data, while PSF-R can reach speckle-level precision, which is essential for high-contrast applications. This underscores the importance of occasionally recording full AO telemetry, both to enable classical PSF-R and to validate the procedures used to derive integrated inputs for Fourier models.

\section{Conclusions}
We have presented a compact, physics-based, and data-calibrated framework for predicting long-exposure, AO-corrected PSFs without relying on full AO telemetry. The approach combines a Fourier-based analytical model with a lightweight neural-network calibrator trained end-to-end with the PSF model, enabling efficient use of limited on-sky training data.

When applied to MUSE-NFM, the approach successfully reproduces polychromatic, field-dependent PSFs in close agreement with on-sky observations. On a held-out validation set of spectrophotometric standard stars, the model achieves a median peak error, $\Delta$SR, of 13.5\% and a relative core FWHM error of 10.9\%. In a crowded $\omega$~Centauri field, the model simultaneously predicts off-axis PSFs for 54 sources with $\Delta$SR $<5$\% and a 4.6\% core FWHM error, enabling clean source separation without per-star PSF fitting.

Most importantly, our results highlight the need to calibrate PSF reconstruction and prediction methods using real on-sky data to achieve the required accuracy for scientific applications. Consequently, PSF models should be conceived from the beginning with calibration capabilities in mind.

Despite the demonstration on MUSE, the framework remains instrument-agnostic: instrument specificity is concentrated in the selected set of reduced-telemetry inputs used to train the small, flexible calibrator, while the analytical core spans multiple AO regimes. This enables this method to be ported to other facilities.

Future work will (i) broaden the training set to fainter and off-axis sources and to moderately crowded fields to strengthen off-axis calibration, (ii) extend the method to other AO regimes and instruments such as SPHERE, and (iii) investigate more real scientific applications of the method.
The implementation is provided in the open-source TipTorch package (see the software availability section).

\section*{Software availability}
\label{sec:software}

The TipTorch package is publicly available under the MIT license: \url{https://github.com/EjjeSynho/TipTorch}

The astro-TIPTOP package is distributed under the MIT License and can be obtained from \url{https://github.com/astro-tiptop/TIPTOP}

The spectrophotometric calibration standard star dataset is available through the ESO archive at \url{https://archive.eso.org/wdb/wdb/cal/qc_products/query?mjd_obs=&instrume==MUSE&pro_catg==DATACUBE_STD&ins_setup==NFM-AO-N_Blue-IR_SCI1.0} 
\begin{acknowledgements}
We want to thank Jo\"el Vernet for his valuable insights into the MUSE instrument and for his overall support of this project. Special thanks go to Sebastian Kamann for providing the $\omega$ Centauri datasets. We also thank Martina Scialpi for testing the early version of the code. We thank Pierre Haguenauer for his help in understanding the MUSE data infrastructure. We would also like to thank Olivier Beltramo-Martin for his assistance in understanding the P3 library's source code and theoretical background. Finally, we extend our thanks to John Pritchard and Fuyan Bian for helping with the MUSE-NFM data reduction pipeline.

This work benefited from the support of the European Southern Observatory, French National Research Agency (ANR) with \emph{WOLF (ANR-18-CE31-0018)}, \emph{APPLY (ANR-19-CE31-0011)} and \emph{LabEx FOCUS (ANR-11-LABX-0013)}; the Programme Investissement Avenir \emph{F-CELT (ANR-21-ESRE-0008)}, the \emph{Action Sp\'ecifique Haute R\'esolution Angulaire (ASHRA)} of CNRS/INSU co-funded by CNES, the \emph{ECOS-CONYCIT} France-Chile cooperation (\emph{C20E02}), the \emph{ORP-H2020} Framework Programme of the European Commission's (Grant number \emph{101004719}), \emph{STIC AmSud (21-STIC-09)}, the R\'egion Sud and the French government under the \emph{France 2030 investment plan}, as part of the \emph{Initiative d'Excellence d'Aix-Marseille Universit\'e A*MIDEX, program number AMX-22-RE-AB-151}.
\end{acknowledgements}

\bibliographystyle{aa}
\bibliography{references}

\begin{appendix}

\section{Theory}
\label{ap:sec:theory}

\subsection{Computing $\text{OTF}_{\text{turb}}$}
\label{ap:sec:OTF_turb}
The OTF term responsible for modeling the effect of imperfect AO correction can be computed from the structure function of the AO residuals, $D_{\varphi}$, which in turn can be calculated using the residuals covariance matrix, $B$, that results from the spatial PSD of the AO residuals, $W$:
\begin{equation}
    \begin{array}{l}
        \vspace{5pt}
        \text{OTF}_{\text{turb}}(\bm{\nu}) = \exp \left( -D_{\varphi}(\bm{\rho}) \, / \, 2 \right),\\
        \vspace{5pt}
        D_\phi(\bm{\rho}) = 2 \left( B(0) - B(\bm{\rho}) \right), \\
        B(\bm{\rho}) = \mathcal{F}^{-1} \{ \, W(\mathbf{k}) \, \}.
    \end{array}
\end{equation}
\noindent
Here,
$\bm{\rho}$ is the spatial shift vector in the pupil plane.

\subsection{Jitter kernel rotation matrix}
Matrix $\mathbf{T}$ is a 2-d rotation matrix defined as
\vspace{-0.25em}
\begin{equation}    
    \mathbf{T}(\theta_J) = 
    \begin{pmatrix}
        \cos \theta_J & -\sin \theta_J \\
        \sin \theta_J &  \cos \theta_J
    \end{pmatrix}.
    \label{ap:eq:jitter_rotation}
\end{equation}

\subsection{PSD error terms}
\label{ap:sec:PSD_errors}

\subsubsection{Fitting error}
\label{ap:sec:PSD_errors:fitting}
According to the Von K\'{a}rm\'{a}n model, the open-loop atmospheric spatial power spectrum is defined as
\begin{equation}
    W_\varphi (\mathbf{k}) = 0.0229 \, r_0^{-\frac{5}{3}} \left(\left\lVert\mathbf{k}\right\rVert^2 + L_0^{-2}\right)^{-\frac{11}{6}}
    \label{ap:eq:von_Karman}
,\end{equation}
where
$r_0$ is Fried's parameter, $L_0$ is the atmospheric outer scale. The piston filter $\widetilde{\Pi}$ in Eq.~\ref{eq:W_fit} results from the inability of differentiating WFSs like Shack-Hartmann WFS to retrieve the average phase over the pupil. It is described by the Bessel function of the first kind defined in the spatial frequency space.

\subsubsection{Wavefront sensor noise error}
\label{ap:sec:PSD_errors:WFS_noise}
The WFS noise variance matrix $\mathbf{C}_b$ is defined as
\begin{equation}
    \mathbf{C}_b = \sigma^2_\eta \ \mathbf{I}_d.
    \label{ap:eq:C_b}
\end{equation}
\noindent
Here, $\mathbf{I}_d$ is the identity matrix, and the total noise variance per WFS subaperture $\sigma^2_\eta$ is composed of three terms:
\begin{equation*}
    \sigma^2_\eta = \left( \frac{\lambda_\text{atm}}{\lambda_\text{GS}} \right)^2 (\sigma_\text{\rm det}^2 + \sigma_\text{\rm ph}^2 + \Delta \sigma^2),
\end{equation*}
\noindent
where
$\sigma_\text{det}^2, \sigma_\text{ph}^2$ are detector readout and photon noise, respectively, defined in Eqs.~(47) and (48) in \cite{Correia:17};
the term $\Delta \sigma^2$ accounts for unexplained noise variance, which is highlighted in Sect.~\ref{sec:inputs_calib};
$\lambda_\text{atm} = 500$ nm is the "atmospheric wavelength," at which all wavelength-dependent quantities, such as $r_0$, are defined;
and $(\lambda_\text{atm} \, / \, \lambda_\text{GS} )^2$ term normalizes $\sigma^2_\eta$ from the WFS wavelength $\lambda_\text{GS}$ to $\lambda_\text{atm}$.

\subsubsection{Spatio-temporal error}
\label{ap:sec:PSD_errors:spatio_temporal}

The projection vectors $\mathbf{P}_{\theta}^L$ and $\mathbf{P}_\alpha^L$ are defined as
\begin{eqnarray*}
    \mathbf{P}_{\theta}^L 
    &=& 
    \biggl(
        \Bigl[
            \exp\!\left( 2 i \pi \, \mathbf{k} \cdot 
                \bigl( \mathbf{h}[l] \, \bm{\theta} - \mathbf{V}[l] \, \Delta t \bigr)
            \right)
        \Bigr]_{\;l,\,1}
    \biggr)_{\,l \,=\, 1}^{N_L}
    \;\in\; \mathbb{C}^{\,N_L \,\times\, 1}, 
    \\[0.25em]
    \mathbf{P}_\alpha^L 
    &=& 
    \biggl(
        \Bigl[
            \exp\!\left( 2 i \pi \, \mathbf{h}[l] \,
                (\bm{\alpha}_i \cdot \mathbf{k})
            \right)
        \Bigr]_{\;l,\,i}
    \biggr)_{\,l \,=\, 1,\;\, i \,=\, 1}^{N_L,\,N_{GS}}
    \;\in\; \mathbb{C}^{\,N_L \,\times\, N_{GS}} ,
\end{eqnarray*}
\noindent with
\begin{eqnarray*}
    \mathbf{w} 
    &=& 
    \biggl(\,
        \Bigl[ \mathbf{w}[l] \Bigr]_{\;l,\,1}
    \biggr)_{\,l \,=\, 1}^{N_L}
    \;\in\; \mathbb{R}^{\,N_L \,\times\, 1}, 
    \\[0.25em]
    \mathbf{V} 
    &=& 
    \biggl(\,
        \Bigl[ \bigl( V_x[l],\, V_y[l] \bigr) \Bigr]
    \biggr)_{\,l \,=\, 1,\,\; i \,=\, 1}^{N_L,\,2}
    \;\in\; \mathbb{R}^{\,N_L \,\times\, 2}, 
    \\[0.25em]
    \mathbf{h} 
    &=& 
    \biggl(\,
        \Bigl[ \mathbf{h}[l] \, \sec\!\bigl(\theta_{\mathrm{zen}}\bigr) \Bigr]_{\;l,\,1}
    \biggr)_{\,l \,=\, 1}^{N_L}
    \;\in\; \mathbb{R}^{\,N_L \,\times\, 1}.
\end{eqnarray*}
\noindent
Here, $N_L$ is the number of atmospheric layers,
$N_{GS}$ is the number of guide stars,
"$\cdot$" is the dot product, 
$\Delta t$ is the WFS integration time,
$\mathbf{w}$ is the vector of the atmospheric layers' weights,
$\mathbf{V}$ is the wind vector per atmospheric layer,
$\mathbf{h}$ is the vector containing the atmospheric layers' altitudes in the observed direction, and
$\theta_{\text{zen}}$ is the zenith angle.
The matrix $\mathbf{C}_\varphi$ weights the solution of tomographic reconstruction by providing priors on the open-loop atmospheric power spectrum per layer and is defined as $\mathbf{C}_\varphi = \mathbf{w} \, \mathbf{I}_d \, W_\varphi.$
Note that $\mathbf{P}_{\theta}^L$ also "absorbs" the spatio-temporal shift of the atmospheric layers.
Square brackets with subscripts, $[\,\cdot\,]_{\,l,i}$, denote individual array elements.
Outer scalable parentheses with sub/superscripts, $(\,\cdot\,)_{\,\cdot}^{\,\cdot}\,$, indicate the assembly of such entries into a vector or matrix over the stated index ranges.
Element access uses small brackets, for example $\mathbf{w}[l]$, to distinguish indexing from the parentheses used for vector/matrix construction.

\subsubsection{Aliasing}
\label{ap:sec:PSD_errors:aliasing}

In Eq.~\ref{ap:sec:PSD_errors:aliasing}, the substitute term $Q(\mathbf{k})$ combines the reconstruction and measurement operators (partially):
\[
    Q(\mathbf{k}) = 2i\pi d \ \mathbf{R}(\mathbf{k}+\mathbf{s}/d) \, \text{sinc}\left(k_m d\right) \,\text{sinc}\left(k_n d\right).
\]
Meanwhile, the term $A_\varphi$ absorbs the spatio-temporal error accumulated over the WFS integration time and AO loop lag:
\vspace{-0.5em}
\[
    \begin{split}
        A_\varphi(\mathbf{k}) &= \widetilde{H}_1 \sum_{l=1}^{N_L} \mathbf{w}[l] \, \text{sinc}\left(k_m V_x[l] \Delta t\right) \ \text{sinc}\left(k_n V_y[l] \Delta t\right) \\
        &\quad\, \times \exp\left(2i\pi t_d (\mathbf{k} + \mathbf{s}/d) \cdot \mathbf{V})\right).
    \end{split}
\]
\noindent
Both terms are defined within an SCAO framework, as elaborated earlier in Sect.~\ref{sec:theory:aliasing}.
Here, $k_m = k_x - m/d$, $k_n = k_y - m/d$ are the spatial frequency vector components that simulate "folding" of spurious spatial frequencies into the domain of the corrected spatial frequencies,
$\widetilde{H}_1 (\mathbf{k})$ is the weighted average of the temporal transfer function over the atmospheric layers introduced in Eq.~14 from \cite{Correia:17},
$\mathbf{R}$ is the vector that models the wavefront reconstruction, which is defined in Eqs.~(24a,\,b) from \cite{Correia:14}.
If the correction is applied in the middle of the integration time $\Delta t$, the total AO loop lag is $t_d = \Delta t /2 + \zeta$, with the last addend being the time lag resulting from WFS readout time and the DM command computation.

\subsubsection{Chromatism error}
\label{ap:chroma}
 The chromatism error arises because shorter-wavelength light undergoes stronger refraction and thus accumulates a larger OPD as they propagate through the atmosphere. As a result, the wavefront correction computed at the WFS wavelength differs from the one required to compensate for the wavefront disturbances at the science wavelength, resulting in imperfect wavefront compensation. 

\subsection{Differential refraction error}
\label{ap:diff_ref}
The differential refraction error arises when the scientific target and GS are observed at some angular separation from the zenith. The atmosphere then acts as a prism, making wavefronts at different $\lambda$ travel slightly different columns of atmospheric turbulence and thus accumulate slightly different wavefront errors.

Then, $\theta_\lambda$ is the apparent angular displacement between two images of the same target observed at $\lambda$ and $\lambda_\mathrm{GS}$, defined as in Eq.~(10) of~\citet{Fusco:06}. In this case, the term $\mathbf{\Theta}_{\mathrm{DR},\,l\:}$ from Eq.~(\ref{eq:diff_ref}) can be denoted as
{\setlength{\jot}{3pt}
\begin{equation}
    \begin{aligned}
        \mathbf{\Theta}_{\mathrm{DR},\,l\:}(\lambda) :=
        2\pi\,h [l] \,\tan(\theta_\lambda)\:\mathbf{z}.
    \end{aligned}
\end{equation}}\unskip
\noindent
Here,
$\mathbf{z} = (1,\,0)^\top$ is the zenith-aligned vector.

\section{Glossary with the variable definitions}
\begin{table}[ht!]
    \centering
    \caption{Glossary of the key variables used in this work.}
    \label{tab:glossary_variables}
    \begin{tabular}{ll}
        \hline\hline
        Parameter & Description \\
        \hline
        \rule{0pt}{2.5ex}
        $r_0(\lambda)$ & Fried parameter $\lambda$. \\
        $L_0$ & Atmospheric turbulence outer scale. \\
        $\tau_0$ & Atmospheric coherence time. \\
        $C_n^2$ & Refractive-index structure constant profile. \\
        $N_L$ & Number of discrete turbulent layers. \\
        $\mathbf{h}$ & Turbulent layers altitudes. \\
        $\mathbf{w}$ & Relative turbulence layer weights. \\
        $\mathbf{V}$ & Turbulence layer wind velocities. \\
        $P(\mathbf{x})$ & Telescope pupil mask. \\
        $\delta x(\lambda),\,\delta y(\lambda)$ & Sub-pixel astrometric shifts in the focal plane. \\
        $J_x(\lambda),\,J_y(\lambda)$ & FWHM of residual TT jitter. \\
        $\theta_J$ & Rotation angle of the TT jitter reference frame. \\
        $F(\lambda)$ & Flux normalization factor. \\
        $b(\lambda)$ & Constant background flux. \\
        $\theta_{\mathrm{zen}}$ & Zenith angle of the observation. \\
        $\bm{\theta}$ & Angular position of the science target. \\
        $\alpha_{\mathrm{pix}}$ & On-sky pixel scale. \\
        $A,\,B$ & Amplitude and background terms of $\Delta W$. \\
        $\alpha_M,\,\beta_M$ & Width and slope parameters of $\Delta W$. \\
        $a_b$ & Scaling coefficient of the static phase bump. \\
        $\Delta\sigma^2$ & Additional noise variance term. \\
        $d_{\mathrm{act}}$ & DM actuator pitch. \\
        $d$ & WFS subaperture size. \\
        $\lambda_{\mathrm{atm}}$ & Atmospheric phase normalization wavelength. \\
        $\lambda_{\mathrm{GS}}$ & Guide-star WFS wavelength. \\
        $N_{\mathrm{GS}}$ & Number of guide stars. \\
        $\bm{\alpha}$ & Angular positions of the guide stars. \\
        $\sigma^2_{\mathrm{det}}$ & WFS detector noise variance.. \\
        $D$ & Telescope primary mirror diameter. \\
        $\Delta t$ & WFS integration time. \\
        $\zeta$ & Effective AO loop delay. \\
        \hline
    \end{tabular}
\end{table}

\section{Calibrator training}
\label{ap:calib_train}
As mentioned earlier, the training of the calibrator $g$ can be formalized as the minimization of the loss function, $\mathcal{L}$:
\begin{equation*}
    \begin{array}{l}
    \min_{\gamma} \frac{1}{N_{I,\,t}} \sum\limits_{k=1}^{N_{I,\,t}} \mathcal{L}\,
    \left(  \,I^{(k)}_\mathrm{data}(\mathbf{p}, \lambda), \: \widetilde{I}^{(k)}(\mathbf{p}, \lambda)  \right), \\
    \widetilde{I}^{(k)}(\mathbf{p}, \lambda) := f \left( \mathbf{p}, \lambda;\; \mathcal{Y}_s \cup \mathcal{Y}_d^{(k)} \oplus g(\mathcal{X}^{(k)}; \gamma) \right).
    \end{array}
\end{equation*}
\noindent
Here, $N_{I,t}=314$ is the number of training samples drawn from on-sky data. Then, for each sample $k$, the loss function is defined as
\begin{equation*}
    \mathcal{L} \,(\,\ldots\,) = 
    \frac{1}{H\,W N_\lambda} \sum^{N_\lambda}_{\lambda}
    \sum_{\mathbf{p} \, \in \, \Omega}
    \omega_1 \left| \Delta^{(k)}(\mathbf{p}, \lambda) \right| \, + \, \omega_2 \left( \Delta^{(k)}(\mathbf{p}, \lambda) \right)^2, \\
\end{equation*}
\begin{equation*}
    \Delta^{(k)} (\mathbf{p}, \lambda) := 
    w(\lambda) \left(\, \widetilde{I}^{(k)}(\mathbf{p}, \lambda) - I^{(k)}_\mathrm{data}(\mathbf{p}, \lambda) \right).
\end{equation*}\unskip
\noindent
The loss combines mean-absolute ($L_1$) and mean-square ($L_2$) terms, weighted by coefficients $\omega_1,\omega_2$ tuned via hyperparameter optimization. The $L_1$ term emphasizes the PSF halo, while $L_2$ focuses on the core energy distribution. The chromatic weighting factor $w(\lambda)$ balances contributions across spectral slices: for MUSE-NFM, "blue" PSFs are more strongly blurred by poorer AO correction, while NIR PSFs are sharper and more core-concentrated.

\section{PSF prediction}
\subsection{FVU definition}
The FVU is a measure of how well a model predicts observed data. It is a normalized error metric, defined as
\begin{equation}
    \text{FVU} = 
    \frac{
        \operatorname{var}\!\left(\,\widetilde{\mathbf{I}} - \mathbf{I}_\text{data}\,\right)
    }{
        \operatorname{var}\!\left(\mathbf{I}_\text{data}\right)
    } \times 100\%.
    \label{ap:eq:FVU}
\end{equation}
A model that perfectly explains the observed data would give FVU $\rightarrow$ 0\%, while FVU > 100\% means that the model is worse than using the mean of the data as a predictor.

\subsection{Calibrated prediction versus direct prediction}
\label{ap:calib_tuned_direct}

Section~\ref{sec:PSF_pred:implementation} describes the calibration scheme implemented via the preceding data-driven component. To quantify its benefit, we consider two simplified PSF-prediction baselines.
\begin{itemize}
    \item Direct prediction. In this configuration, all model-related inputs available in $\mathcal{X}$ are provided directly to the model, while $F(\lambda)=1$ at all wavelengths and $\Delta\sigma^{2}=A=B=\alpha_M=a_b=0$. The TT jitter is assumed to be achromatic and is computed by applying Eqs.~(47)–(48) of \cite{Correia:17} to the $2 \times 2$ subaperture WFS. In other words, here the PSF model is used purely analytically without any calibration or error "absorption."
    
    \item Tuned–direct prediction. Starting from the direct configuration, we apply a minimal, data-driven correction by optimizing global median values for the variables in $\mathcal{Y}_c$ (excluding $r_0$ and $\mathbf{w}$) to minimize the discrepancy between the PSF model and the training subset of $\mathcal{D}$. These optimized quantities act as static offsets in model inputs, rather than functional dependencies on wavelength or field position.
\end{itemize}

\begin{figure}[h!]
    \centering
    \includegraphics[width=0.485\textwidth]{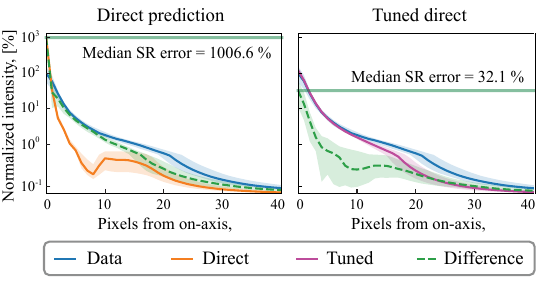}
    \caption{Radial profiles for the direct (left) and tuned-direct (right) baselines, computed for white-light PSFs.}
    \label{ap:fig:profiles_direct_tuned}
\end{figure}

In both cases, the photometric bias ($b$) and astrometric shifts $(\delta x,\delta y)$ are obtained by image-plane fitting for each sample. Performance on the validation set for the direct and tuned–direct approaches is summarized in Fig.~\ref{ap:fig:profiles_direct_tuned} and Table~\ref{tab:direct_tuned_FVU}.

\begin{table}[ht!]
    \centering
    \renewcommand{\arraystretch}{1.2}
    \setlength{\tabcolsep}{10pt}
    \caption{Median validation metrics for different prediction approaches.}
    \begin{tabular}{lccc}
        \hline\hline
        Metric & Direct & Tuned–direct & Calibrated \\ \hline
        $\Delta\mathrm{SR}$, [\%] & 1006.6 & 32.1 & 13.5  \\
        $\Delta \mathrm{FWHM}_{\mathrm{rel}}$, [\%] & 80.0 & 20.2 & 10.9 \\
        FVU [\%] & 8610 & 17.4 & 3.2 \\ \hline
    \end{tabular}
    \label{tab:direct_tuned_FVU}
\end{table}
Relative to the calibrated prediction reported in Sect.~\ref{sec:results:STD}, both simplified baselines yield significantly poorer prediction accuracy, underscoring the necessity of the data-driven calibration.

Two main factors account for the large errors obtained with direct prediction. First, MUSE PSFs are under-sampled and exhibit a steep exponential profile. Consequently, even small inaccuracies in the PSF model parameters translate into significant errors in the central pixel intensity, which in turn strongly affect the SR estimate. These peak-intensity errors also highlight the need for accurate sub-pixel astrometric fitting. Second, an adequate description of TT jitter is crucial, as it largely determines the FWHM of the PSF. In the direct approach, the TT wavefront error is derived from LO WFS noise, which underestimates its power. As a result, the predicted PSFs appear significantly sharper than observed in practice, reducing the model's ability to reproduce the overall PSF morphology and thereby increasing the FVU error.

\end{appendix}

\end{document}